# The Pharmaceutical Price Regulation Crisis: Implications on Antidepressant Access for Low-Income Americans

Nicole Hodrosky*[1], Gabriel Cacho*[2], Faiza Ahme[3], Rohana Mudireddy[4], Yapin Wen[5], Kymora Nembhard[6], Michael Yan[7]

*These authors contributed equally to this work.

**Affiliations:**

1. Boston University, Boston, MA
2. Hillsborough Community College, Ruskin, FL
3. Fordham University, New York, NY
4. Independence High School, San Jose, CA
5. Brooklyn Technical High School, Brooklyn, NY
6. Howard University, Washington, D.C.
7. University of California, Berkeley (UC Berkeley), Berkeley, CA, USA

**Corresponding Author:**
Gabriel Cacho
Hillsborough Community College, Ruskin, FL

**Abstract**

Depression affects more than 280 million people worldwide, with poorer communities having disproportionate burden as well as barriers to treatment. This study examines the role of pharmacy pricing caps in access to antidepressants among poorer Americans through bibliometric analysis of the 100 most cited articles on antidepressant pricing and access in the Web of Science Core Collection. We used tools like Bibliometrix and VOSviewer to visualize publication trends, dominant contributors, thematic clusters, and citation networks in the literature. Findings highlight intransigent inequalities in access to antidepressants based on astronomically high drug pricing as well as systemic inequalities against racial and ethnic minorities in particular. Branded antidepressant high prices are associated with low initiation of therapy as well as regimen compliance, heightened mental illness outcomes, as well as increased health utilization. This work uncovers critical gaps in the literature and demands immediate policy action to make antidepressants affordable as well as appropriately accessible to marginalized communities.

**Introduction**

Depressive disorders impact more than 280 million people worldwide, affecting individuals' daily lives whether in social interactions or basic needs.[1] Symptoms often include: feelings of sadness or hopelessness, loss of interest in normal activities, sleep disturbances, lack of energy, appetite changes, anxiety, slowed thinking, trouble concentrating, or suicidal thoughts.[2] Low-income individuals are nearly 2.5 times more likely to experience depression compared to those at or above the federal poverty level.[3]

In terms of treatment, generic antidepressants include SSRIs (fluoxetine and sertraline), SNRIs (venlafaxine and duloxetine), and TCAs (amitriptyline). They work by stabilizing brain chemicals such as serotonin and norepinephrine to regulate mood and alleviate depression symptoms.[4] While antidepressants, such as SSRIs, SNRIs, and TCAs, are commonly prescribed and beneficial for many individuals, their actual effectiveness can vary significantly depending on the individual and their specific condition.[5] The effectiveness of SSRIs' in terms of treating depression is contentious due to the disparity between clinical trial findings and those obtained in routine practice. Retrospective data generally offer contradictory evidence regarding their cost-effectiveness and actual impact on patient health.[6]

Between the early 1990s and 2000s, the rate of antidepressant drug treatment in the US more than quadrupled. The *Lancet* (2016) conducted a large analysis and found that 60-80% of patients respond to antidepressants. While this increase occurred across all demographic groups, disparities persisted: younger adults, men, and racial/ethnic minorities continued to receive treatment at lower rates than their counterparts.[7] The antidepressant Prozac is $150.10 for a 30 day supply compared to the older version, amitriptyline, which is roughly $3 for the generic version.[5] The higher cost-sharing for branded antidepressants is associated with lower probabilities of treatment initiation or continuation among patients. Racial and ethnic minority patients are also less likely than White patients to receive optimal treatments for depression and anxiety after entering care, independent of cost factors, highlighting broader systemic disparities in mental health treatment access and quality. These inequalities are most directly observed within Black and Hispanic patients with depression, who are less likely to receive mental health treatment, including antidepressant medications, in comparison to non-Hispanic Whites.[8]

These medications become prohibitively inaccessible due to their price, inability to get insurance coverage, and therefore, deprive lower socioeconomic status groups from acquiring much-needed medications. This contributes to the remaining percentage of patients who either have a lower response rate or none at all.[9] Moreover, financial burden due to high medication costs compromises appropriate medication use, adherence, and the ability to continue or fill the prescription. Lack of continued treatment can heighten symptoms of depression, but also perpetuate a cycle of poorly treated mental health conditions that negatively impact the well-being of under-resourced populations. This can result in increased utilization of other healthcare services, such as emergency room visits and hospitalizations, which adds to the overall cost of healthcare.[3,10]

Bibliometric analysis was initially carried out using the Web of Science Core Collection database to analyze how unregulated pharmaceutical pricing may impact antidepressant access. 100 most-cited articles on the subject of antidepressant pricing and access were generated from the Web of Science Core Collection, hence providing insight into the most updated research regarding this topic. The Bibliometrix platform analyzed articles on pharmaceutical prices, mental health, and other related areas. It involves statistical methods to study publication patterns, citation trends, and inter-relationships between themes in studies, authors, and institutions by identifying principal themes in studies, seminal studies, and principal contributors. By applying bibliometric analysis, trends and gaps in the literature on pharmaceutical pricing and antidepressant access were revealed. Network analysis and citation mapping then traced back how such problems emerged, the geographic distribution across studies, and gaps in studies in areas that require policy action or additional research. The analysis also identified frequently cited papers and influential authors.

## Methods

To analyze low-income Americans' accessibility to antidepressants alongside pharmaceutical pricing trends, a bibliometric analysis was conducted. The search was performed in the Web of Science Core Collection database from its inception through January 2025. Multiple keyword combinations were utilized to generate a broad range of relevant literature, including key terms such as "pharmaceutical pricing," "antidepressants," "cost," "access," "depression," "medication," and "low-income." Articles were manually reviewed for relevance after each search, then gathered into a combined list, yielding 220 peer-reviewed articles. The articles were sorted in descending order of citation count, and put through additional stages of manual review, down to 154 articles to the final yield of 100 top-cited articles. Inclusion criteria for the review required articles with a primary focus on the junction of pharmaceutical pricing and access to antidepressant medication, with particular emphasis on low-income populations. Primary research and review articles published with full text in peer-reviewed journals indexed by the Web of Science were also considered for eligibility. Among these eligible studies included 82 articles and 18 review journals. Exclusion criteria included articles not centered around pharmaceutical pricing or antidepressant access, articles not available in full text, abstracts or conference posters, studies published in low-impact factor journals, or those with a poor peer-review standard. The final selection ensured that the included articles provided relevant, high-quality data primarily from the United States for wider implications of antidepressant pricing on public health.

After scoping for relevant articles, several metrics for each analysis were collated: article title, publication year, first and senior authors, author institutions, journal name, keywords, total and annual average citations, and country of origin. Descriptive statistics were used to analyze citation trends, publication patterns, and geographic distribution. Additionally, network analysis of the author relationships and keyword co-occurrence was conducted to identify main research themes and collaborations across the research field. The applied tool for networks and mapping includes VOSviewer, Version 1.6.16 and Bibliometrix-a package for R programming. These tools visualized the relationships among the key research topics and author affiliations, highlighting the highly influential studies and debates that persist in the area. Inclusion criteria for bibliometric analysis were the 100 top-cited articles published in peer-reviewed journals, related to mental health, depression, antidepressant access, and/or pharmaceutical pricing. Each article had a citation count of at least 50 and was otherwise excluded, especially those not pertaining to the focus of the analysis or lacking full data on author affiliations. Further research is needed to address issues of accessibility and affordability, particularly regarding antidepressant medications among low-income populations. By identifying the top 100 cited articles (Figures 1,3, and 4), this analysis presents a comprehensive landscape of the current state of research on antidepressant pricing, contributing valuable insights to inform future policy discussions and academic inquiries.

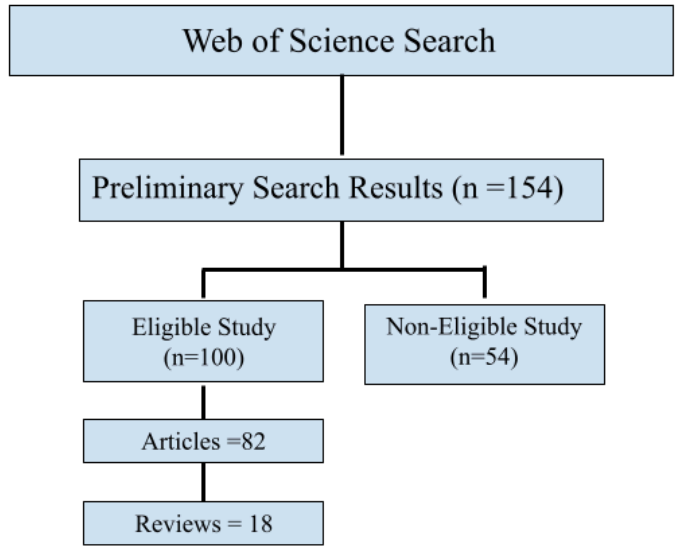

*Figure 1.* Flow diagram of the literature search process outlining the number of studies that were identified, included, and excluded at each stage.

## Results:

**Top 100 most-cited articles**

Analysis of the 100 eligible studies on pharmaceutical price regulation yielded several insights. The article *"Addressing the burden of mental, neurological, and substance use disorders: key messages from Disease Control Priorities"* (Patel, Chisholm, Parikh, et al.), published in *The Lancet*, had the highest citation count (n=538).[7] Harvard University was the most frequently associated institution for senior authors (n=15), followed by the University of Michigan (n=14) and Johns Hopkins University (n=13). The most recurring journals in the dataset included *Journal of Mental Health Policy and Economics, Psychiatric Services, General Hospital Psychiatry, and JAMA: The Journal of the American Medical Association* (each n=4). Frequently mentioned keywords included "mental-health" (349 instances), "pharmaceutical" (220 instances), and "disparities" (150 instances).

**Senior Author Productivity**

There were 6 top senior authors who had multiple papers cited amongst the 100 papers generated. The most productive author was Ell, K, publishing 4 papers, meanwhile the remaining 5 authors each respectively published 2 papers .

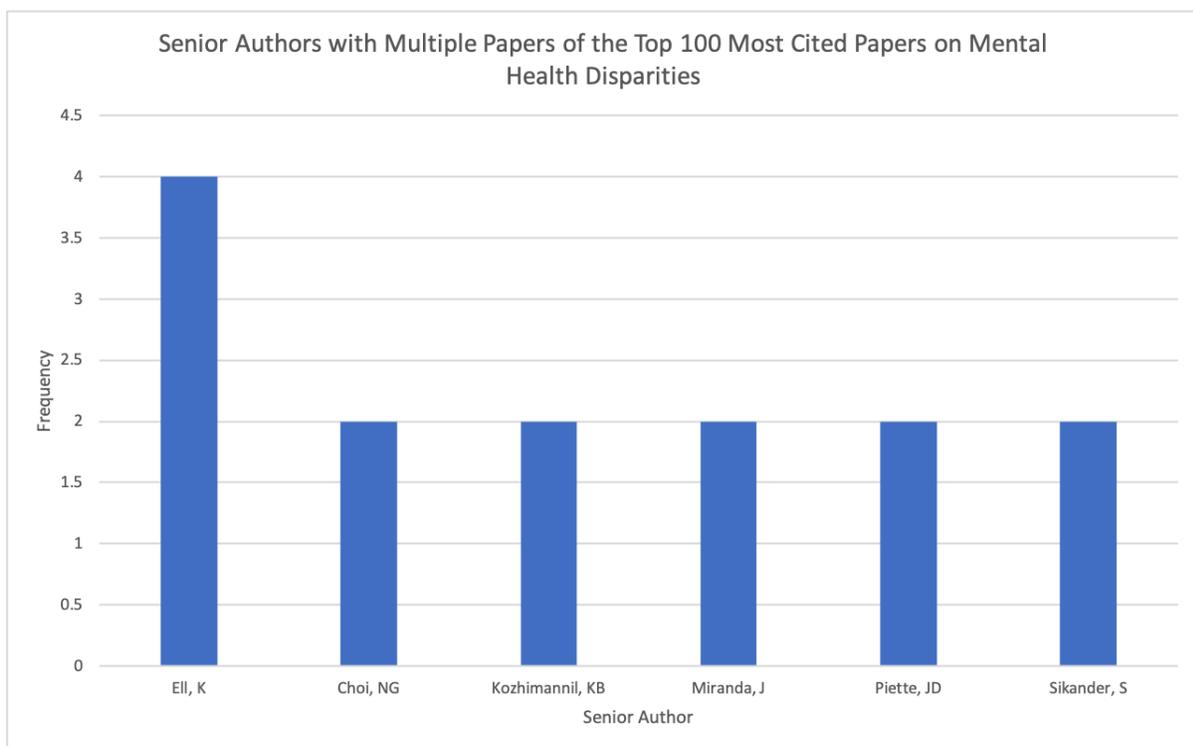

*Figure 2:* Senior Author Productivity, measured by the number of publications per author.

**Institutions of Publication**

13 different institutions were included, each with at least 3 publications. Amongst the senior authors, the leading institution is Harvard University with 15 papers, followed by University of Michigan with 14 papers, and Johns Hopkins University with 13 papers.

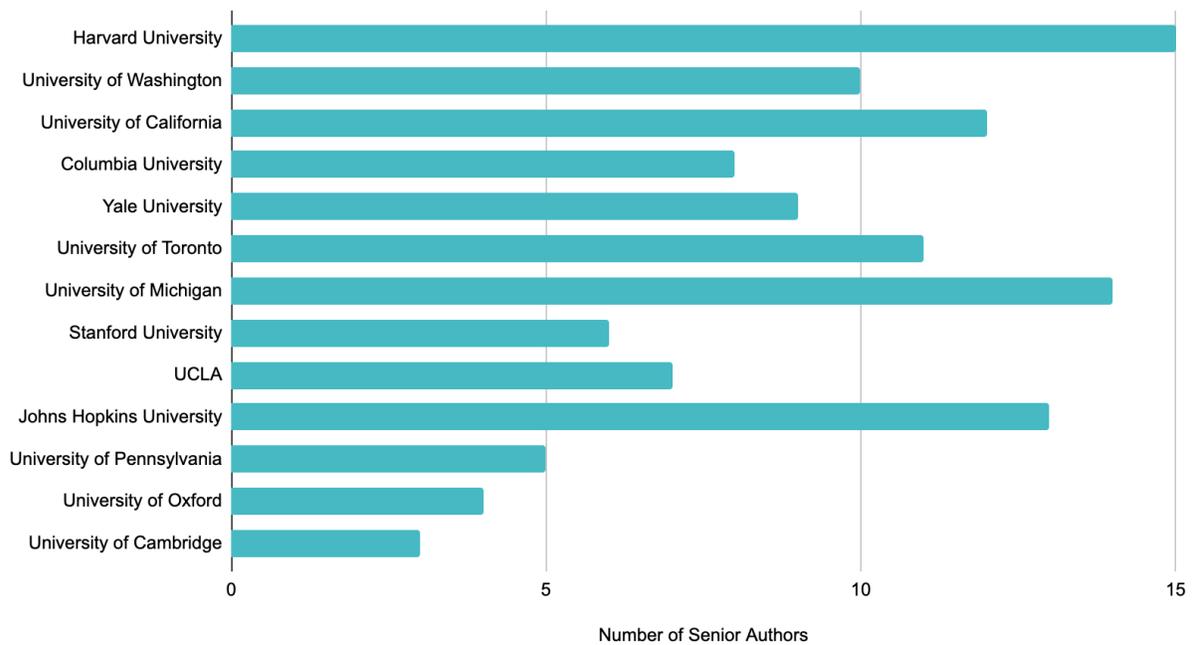

*Figure 2.* Institutions of Publication showing the distribution of publications across 13 different institutions.

**Most Recurring Journals**

25 different journals were listed among the top 100 cited articles, where all had a frequency of at least 1. The most recurring journals were the Lancet, Medical Care, Social Psychiatry and

Psychiatric Epidemiology, and Cognitive and Behavioral Practice.

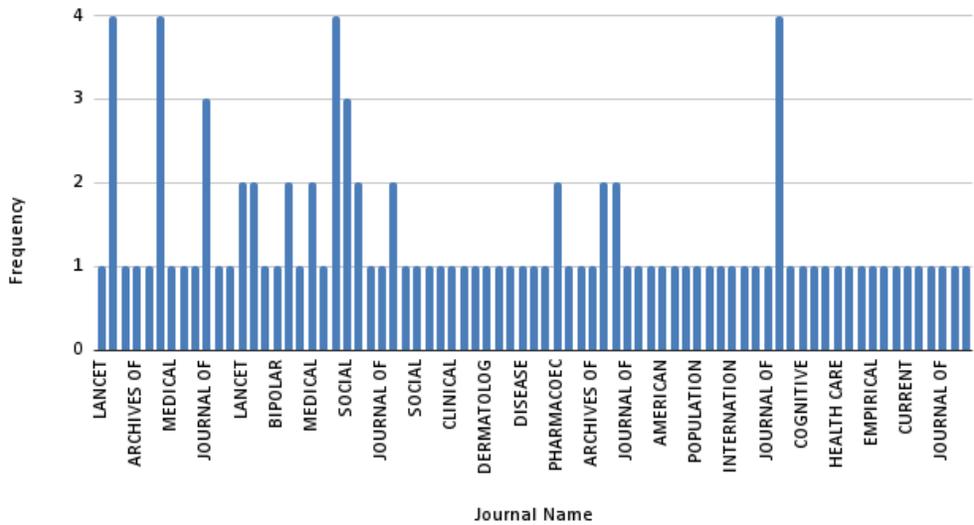

***Figure 3:*** *Most Recurring Journals showing the distribution of top-cited articles across 25 different journals, with a focus on the most frequently represented publications.*

**Frequently Mentioned Keywords**

The top 100 articles generated keywords of mental-health, pharmaceutical, disparities, drug, recovery, low-income, and healthcare. Mental-health was listed roughly 350 times within the articles, followed by pharmaceutical at roughly 220 times, and disparities at 150 instances.

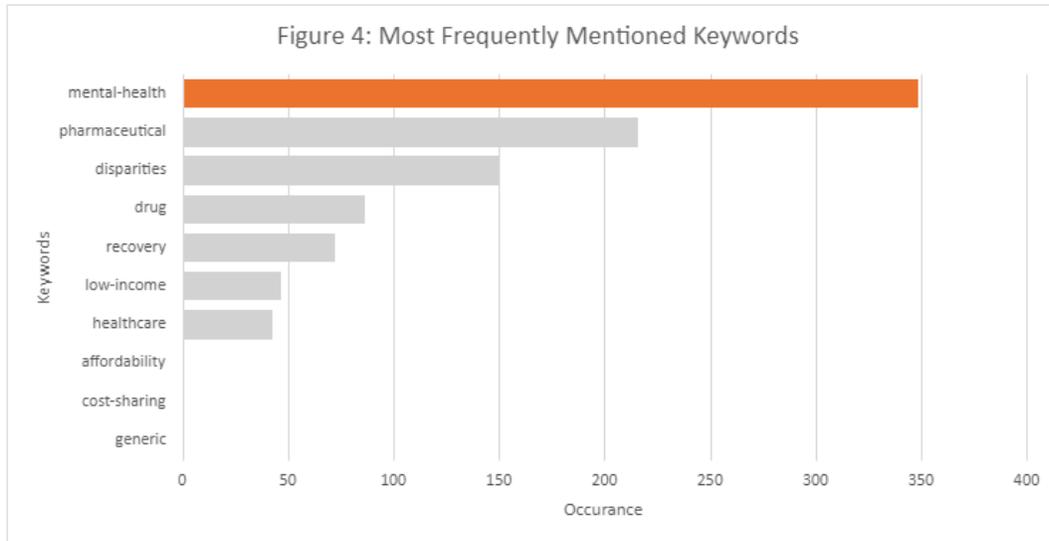

*Figure 4:* Frequently Mentioned Keywords highlighting the most common terms across the top 100 articles

*Table 1. Top 26 Most Cited Articles Regarding Pharmaceutical Price Regulations*

| First Author | Senior Author | Article Title | Journal Title | Total Citations |
|---|---|---|---|---|
| Patel, Vikram | Whiteford, Harvey | Addressing the burden of mental, neurological, and | LANCET | 528 |
| Goldman, DP | Teutsch, SM | Pharmacy benefits and the use of drugs by the chr | JAMA-JOURNAL OF THE AMERICAN MEDICAL ASSOCIATION | 434 |
| Miranda, J | Belin, T | Treating depression in predominantly low-income | JAMA-JOURNAL OF THE AMERICAN MEDICAL ASSOCIATION | 414 |
| Sommers, Benjamin D. | Epstein, Arnold M. | Changes in Utilization and Health Among Low-Inco | JAMA INTERNAL MEDICINE | 378 |
| Piette, JD | Kerr, EA | The role of patient-physician trust in moderating | ARCHIVES OF INTERNAL MEDICINE | 266 |
| Edlund, MJ | Kessler, RC | Dropping out of mental health treatment: Patterns | AMERICAN JOURNAL OF PSYCHIATRY | 266 |
| Givens, Jane L. | Cooper, Lisa A. | Ethnicity and preferences for depression treatmen | GENERAL HOSPITAL PSYCHIATRY | 251 |
| Heisler, M | Piette, JD | The health effects of restricting prescription medic | MEDICAL CARE | 235 |
| STURM, R | WELLS, KB | HOW CAN CARE FOR DEPRESSION BECOME MORE | JAMA-JOURNAL OF THE AMERICAN MEDICAL ASSOCIATION | 234 |
| Ell, K | Russell, C | Depression, correlates of depression, and receipt c | JOURNAL OF CLINICAL ONCOLOGY | 231 |
| Donker, T. | Christensen, H. | Economic evaluations of Internet interventions for | PSYCHOLOGICAL MEDICINE | 209 |
| Katon, Wayne | Jones, Loretta | Collaborative depression care: history, evolution a | GENERAL HOSPITAL PSYCHIATRY | 189 |
| Silverman, Julie | Nelson, Karin | The Relationship Between Food Insecurity and Dep | JOURNAL OF GENERAL INTERNAL MEDICINE | 138 |
| Danzon, PM | Chao, LW | Does regulation drive out competition in pharmace | JOURNAL OF LAW & ECONOMICS | 129 |
| Aguilera, Adrian | Munoz, Ricardo F. | Text Messaging as an Adjunct to CBT in Low-Incom | PROFESSIONAL PSYCHOLOGY-RESEARCH AND PRACTICE | 116 |
| Santos, Andreia Costa | Bull, Fiona C. | The cost of inaction on physical inactivity to public | LANCET GLOBAL HEALTH | 113 |
| Sikander, Siham | Rahman, Atif | Delivering the Thinking Healthy Programme for per | LANCET PSYCHIATRY | 111 |
| Choi, Namkee G. | Kunik, Mark E. | SIX-MONTH POSTINTERVENTION DEPRESSION AND | DEPRESSION AND ANXIETY | 111 |
| Bryant-Comstock, L | Devercelli, G | Health care utilization and costs among privately in | BIPOLAR DISORDERS | 111 |
| Reynolds, Charles F., III | Albert, Steven M. | Early Intervention to Reduce the Global Health and | ANNUAL REVIEW OF PUBLIC HEALTH, VOL 33 | 104 |
| Russell, JM | Rush, AJ | The cost consequences of treatment-resistant depr | JOURNAL OF CLINICAL PSYCHIATRY | 104 |
| Vega, William A. | Ang, Alfonso | Addressing stigma of depression in Latino primary | GENERAL HOSPITAL PSYCHIATRY | 102 |
| Le Cook, Benjamin | Sanchez, Maria Jose | A Review of Mental Health and Mental Health Care | MEDICAL CARE RESEARCH AND REVIEW | 99 |
| Miranda, J | Revicki, D | One-year outcomes of a randomized clinical trial tr | JOURNAL OF CONSULTING AND CLINICAL PSYCHOLOGY | 87 |
| Interian, Alejandro | Vega, William A. | Stigma and Depression Treatment Utilization Amor | PSYCHIATRIC SERVICES | 84 |
| Flores, Elaine C. | Simms, Victoria | Mental health impact of social capital intervention | SOCIAL PSYCHIATRY AND PSYCHIATRIC EPIDEMIOLOGY | 81 |

**Discussion:**

Our findings bring to the forefront the pivotal contribution of drug price in influencing access to mental health medication among the poor. The bibliometric synthesis gathered from the Web of Science Core Collection (WoSCC) established consensus in the literature that increasing drug prices are a primary hindrance to treatment, exacerbating mental health disparities. The economic burden of mental health disorders is real and growing, with research estimating that total global economic output lost on mental, neurological, and substance abuse disorders was $8.5 trillion in 2010. This is expected to double by 2023 if substantial interventions are not implemented.[7] A study from the early 2000s titled *Pharmacy Benefits and the Use of Drugs by the Chronically Ill,* indicated that economic obstacles generally dissuade patients from adhering to prescribed antidepressant treatment, further complicating their conditions and jeopardizing even worse mental health outcomes like suicide.[4] These findings highlight the future need for financial investment in mental healthcare policies and affordability strategies to mitigate long-term economic consequences.

The prevalence of high-impact institutions such as Harvard University, the University of Michigan, and Johns Hopkins University in this research field suggests that leading research institutions have prioritized this (Figure 2). However, evidence of comparatively sparse research from the 2008 study, *New Evidence Regarding Racial and Ethnic Disparities in Mental Health: Policy Implications,* highlights the intersection of uncontrolled drug prices and mental illness disparities suggesting a gap in the literature.[11] Future studies would be well advised to measure the direct impact of cost-related nonadherence to mental illness medication, particularly among the underserved.

Our keyword analysis indicates increasing research interest in inequalities and economic barriers, since awareness of systemic problems behind the impediments to equal healthcare access continuously increases (Figure 4). Policy-oriented research continues to be limited, with studies such as *The Role of Prices in Excess US Health Spending,* emphasizing that government action through price control policy or higher subsidies can reduce patients' expenses.[12] Alternatively, research from the 2020 study, *Cost-related Medication Nonadherence and Its Risk Factors Among Medicare Beneficiaries,* shows a prospective system could be created that tracks patients who seek and receive coverage from PAPs and evaluates their subsequent patterns of medication use, including that which is provided by public programs.[13] Other alternatives, including encouragement of generic drugs, telemedicine-based psychiatric care, and nongovernmental pharmaceutical programs, may improve access to care.[5] In addition, while this analysis focuses primarily on the economic aspects of mental health care, other pressing issues deserve further study. These include gaps in insurance coverage, shortages of mental health and general healthcare providers, and the ongoing stigma surrounding mental health treatment. These challenges go beyond pricing and reflect deeper structural and social barriers. Addressing them will require cross-disciplinary solutions from economics, public health, psychology, and policy analysis. Exploring these broader implications is essential for developing long-term, effective strategies toward improving mental health care access and outcomes.

The financial burden of mental health treatment is particularly evident in low-income and lower/middle-income countries. According to the *Lancet* article, *Addressing the Burden of Mental, Neurological, and Substance Use Disorders: Key Messages from Disease Control Priorities,* increasing cost-effective mental health interventions would cost only $3-4 per person per year. However, due to sustained underfunding less than 1% of total health-related

government spending in these countries is allocated to mental health services.[7] Minimal proportion of investment goes towards mental health, as the majority of these investments are allocated toward hospital construction, emergency treatment, and the development of medicines. The lack of attention toward mental health prevention and treatment turns patients away from assistance, accompanied by long waiting lists and a lack of mental health specialists. They also concern themselves with how government funds that do exist are utilized. Most of it gets used on expensive hospitals and drugs, as opposed to mental health treatment in the area. In the meantime, drugs aren't strictly regulated in certain regions. That tends to give birth to problems like doctors over-prescribing and drugs growing expensive, and unequal access to treatment. That raises the question of whether or not companies have a preference for helping people or making a profit.

These problems show how healthcare spending needs to be fairer and more transparent. Without substantial government financial protective measures due to lack of insurance and funding, many individuals are forced to pay themselves, causing economic disparities and limiting access to care. Recent findings from 2023 show a shift towards universal public financing could lead to better equitable distribution of mental health resources across groups with different levels. But, achieving this requires financial resources and targeted policy efforts to address systematic barrier shortage of mental healthcare providers, weak governance, and stigma surrounding mental health.[7] To improve global mental health outcomes, it is essential to expand financial support systems and ensure sustainable resource allocation.

In this analysis, we observed the importance of understanding the details of pharmaceutical pricing and its impact on the affordability of antidepressant medications for the underprivileged segments of the population in the United States. The age-old financial barriers of mental health care need to be made urgently by policymakers, researchers, and healthcare professionals too. By identifying the top 100 cited articles in this field, our analysis provides valuable insights into the ongoing research landscape and highlights the necessity for continued exploration of potential interventions. A 2004 analysis that analyzed pharmacy benefits showed that increasing medication costs have posed long standing challenges for the chronically ill, emphasizing the importance of continued research in this area.[4] According to a 2005 analysis, disparities in access to mental health care persist, especially for impoverished minority women who experience a higher burden from depression than white women and are less likely to receive appropriate care.[14] The sample in our analysis was predominantly poor, with more than half of the women living at or below the federal poverty guidelines. This aligns with findings from the *CDC's* 2014 report, *Depression in the U.S. Household Population, 2009-2012,* which found that individuals living below the poverty level are twice as likely to experience depression and significantly less likely to receive any care, with only 35% of those with severe symptoms reporting contact with a mental health professional in the past year.[15] Effective interventions from studies in 2003, included medications and psychotherapy, significantly reduced depressive symptoms and improved social functioning among these women compared with community referral.[14] Future efforts should emphasize expanding affordability through pharmaceutical price regulations and alternative cost-reduction strategies. Ensuring equitable access to antidepressant medications remains a critical public health priority that requires immediate and sustained action. Results indicate that constant investigation and policy development should be encouraged in the field of accessing antidepressant medications at an affordable price by people of low incomes.

## Limitations

However, this analysis has several limitations, including the lack of generalizability to low and middle-income countries and limited data on the long-term impact of pharmaceutical price regulations on medication access. This analysis, while making useful observations, has some underlying flaws which can be rectified in future research.

Firstly, its observations relate mainly to high-income countries, and in particular, the United States, where the healthcare system is largely privatized and insurance varies widely across populations, resulting in inconsistent access to medications. Although programs like Medicaid and Medicare aim to support low-income and elderly individuals, gaps in coverage and strict eligibility requirements can leave many without affordable access to essential mental health treatments. This contrasts significantly with low and middle-income countries (LMICs), where underfunded healthcare systems present barriers to care. In those countries, healthcare is further restricted, and mental health frameworks remain underfinanced, so drug pricing effects and non-adherence effects can be multifarious. Research is necessary in undertaking studies to determine ways healthcare disparities in those countries, economically and structurally, impact mental health consequences. Also, pharmaceutical pricing regulation data and drug expenditure containment measures have limited longitudinal data to account for direct influences on drug availability and patient outcomes. Much existing literature in this research field uses observations within limited time horizons, and additional longitudinal research is essential to determine whether and for how long controls and subsidies, and other financial measures can sustainably support better mental health medication availability. Lastly, in this review, frequently cited papers were utilized, which imposes an unavoidable bias towards highly cited papers in academic circles. Highly cited papers do warrant use, yet they do not include all views, and opinions of underrepresented, marginalized communities' views may not be fully captured. Future research should include diverse papers from various sources in an attempt at an improved reflection of drug pricing's impact upon mental health worldwide.

Furthermore, the research primarily discussed availability of medication in relation to affordability and adherence rather than in relation to overall access to pharmacies or health centers. In rural and underserved communities, even when medication exists, patients do not have easy access to mental health professionals, clinicians, and pharmacies due to logistical barriers, for instance, transit difficulties or lack of healthcare infrastructure. This also contributes further to exacerbating gaps in accessing care and should be explored in future research so that an overall better understanding of barriers to care in patients can be ascertained.

Analysis also discusses almost solely economics in accessing mental health medication and does not include consideration for the impact of stigma and culture in discouraging patients from seeking care and adhering to medicines. Cultural stigma revolving mental illness in minority and low-income communities can significantly hinder individuals from seeking care. Moreover, gender disparities play a critical role, as women are more likely to be diagnosed with depression but they are often dismissed whereas men, due to social expectations and masculinity, avoid seeking help. These overlapping factors, like culture, race, and gender, create barriers that future studies must address in further detail. In second place, although medication is key in so many patients with mental disease, this review discusses pharmaceuticals virtually solely as interventions, perhaps at the expense of other efficacious interventions, for instance, psychotherapy, community-based programs, and peer-based support networks. A review of all mental health interventions available would provide a better balance in terms of possibilities for mental health disparities. Another problem is this research does not adequately address the variation in health insurance plan and regimes of various regions and socioeconomic strata. In the U.S., for instance, mental health medication varies in terms of insurance status, entitlement, and state level statute, and this can affect adherence to medication and access. Finally, although this review discusses antidepressants, other classes of medication for mental health medication, for instance, antipsychotics, mood stabilizers, and anxiolytics, receive short shrift. Every one of these medicines has its own pricing structure, barrier for access, and therapeutic effect which can affect adherence and mental state. Addition of such medicines would augment our understanding in terms of mental health care access issues.

## **Conclusion:**

Our research highlights that the high costs of antidepressants remain a significant barrier for low-income patients, exposing a critical flaw in the current system. Untreated depression due to these financial barriers exacerbates mental health issues, leading to increased healthcare costs, higher illness rates, and elevated mortality. Through bibliometric analysis, key trends, influential studies, gaps in the literature, and limitations arose regarding pharmaceutical pricing and access to antidepressant medications. Analysis indicates a need for further research and policy change in this area, particularly for marginalized and underserved populations. While our analysis does not propose specific interventions, the patterns observed underscore the necessity of continued investigation into achieving pricing and healthcare affordability to mitigate persistent disparities in depression treatment. Addressing the unregulated pricing crisis in pharmaceuticals will improve access to antidepressant medications and lead to progressive mental health outcomes among low-income Americans. The approach may include comprehensive policy reform in drug pricing regulation, development of low-cost treatment options, and increased access to mental health services that will reduce financial barriers to necessary medications.